\documentclass[conference]{IEEEtran}
\usepackage[utf8]{inputenc}
\usepackage{graphicx}
\usepackage{amssymb}
\usepackage{amsmath}
\usepackage{makecell}
\usepackage{multicol}
\usepackage{authblk}
\usepackage{float}
\usepackage{cite}
\begin{document}

\title{Unified Resource Allocation Framework for the Edge Intelligence-Enabled Metaverse\vspace*{-10mm}}

\author{Wei Chong Ng$^{1,2}$, Wei Yang Bryan Lim$^{1,2}$, Jer Shyuan Ng$^{1,2}$, Zehui Xiong$^{3}$, Dusit Niyato$^{4}$ and Chunyan Miao$^{4,5}$\\
$^1$Alibaba Group~$^2$Alibaba-NTU Joint Research Institute~
$^3$Singapore University of Technology and Design\\
$^4$School of Computer Science and Engineering, Nanyang Technological University, Singapore \\$^5$Joint NTU-UBC Research Centre of Excellence in
Active Living for the Elderly (LILY)\\
\vspace*{-5mm}}
\maketitle

\begin{abstract}
Dubbed as the next-generation Internet, the metaverse is a virtual world that allows users to interact with each other or objects in real-time using their avatars. The metaverse is envisioned to support novel ecosystems of service provision in an immersive environment brought about by an intersection of the virtual and physical worlds. The native AI systems in metaverse will personalized user experience over time and shape the experience in a scalable, seamless, and synchronous way. However, the metaverse is characterized by diverse resource types amid a highly dynamic demand environment. In this paper, we propose the case study of virtual education in the metaverse and address the unified resource allocation problem amid stochastic user demand. We propose a stochastic optimal resource allocation scheme (SORAS) based on stochastic integer programming with the objective of minimizing the cost of the virtual service provider. The simulation results show that SORAS can minimize the cost of the virtual service provider while accounting for the users' demands uncertainty.
\end{abstract}

\section{Introduction}
The recent COVID-19 pandemic~\cite{bick2020work} has driven the rise in adoption of the online virtual environment as a viable alternative for a growing range of shared human experiences. For example, UC Berkeley held its graduation ceremony in Minecraft\footnote{https://news.berkeley.edu/2020/05/16/watch-blockeley-uc-berkeleys-online-minecraft-commencement/}, that was originally created to be a gaming platform. Together with the advancement of other enabling technologies ranging from the advent of $5$G to blockchain to Virtual Reality (VR)/Augmented Reality (AR), the conditions towards the development of the \textit{metaverse}, also known as the next-generation internet, have been gradually fulfilled.

As described by venture capitalist Matthew Ball \cite{ball}, the metaverse is an embodied version of the Internet. Similar to sci-fi films such as \textit{Ready Player One}~\cite{readyplayer}, users can leverage VR/AR technologies to navigate freely  within a virtual environment with their customized avatars. Beyond gaming functions, the metaverse can support novel ecosystems of  service provisions that will blur the lines between the physical and virtual worlds. For example, the American rapper Lil Nas X held his concert online on the Roblox platform\footnote{https://techcrunch.com/2020/11/10/roblox-to-host-its-first-virtual-concert-featuring-lil-nas-x/}, whereas the Facebook Infinite Office enables users to collaborate with their colleagues in an immersive online environment.

\subsection{Overview of Metaverse and Virtual Education Case Study}
The metaverse can be considered to  follow a three-layer architecture~\cite{duan2021metaverse}. Firstly, the \textit{physical} layer consists of all the hardware to support the operational functions of the metaverse, i.e., computation, communication, and storage. A robust physical layer is of utmost importance to ensure the scalable and ubiquitous access to the metaverse. Secondly, the \textit{virtual} layer provides a parallel living world in which the users' avatars can interact with each other or with other objects. For immersivity, the virtual layer should also capture and reflect the real-time data and analytics of the real world using technologies such as edge intellgience empowered digital twins~\cite{el2018digital}. Finally, the \textit{interaction} layer serves as a bridge to connect the users in the physical world to the virtual world. The users can upload their inputs in the physical world that is eventually translated  into specific actions in the virtual world. To enhance service delivery, the wealth of data collected from the users can be leveraged to train the Artificial Intelligence (AI) system to deliver highly personalized services to users.

Similar to the mobile internet, the Quality of Service (QoS) provided by the metaverse is a critical factor that will determine its successful adoption. Users will expect the metaverse to be scalable, seamless, and synchronous. In addition, the AI system in the metaverse can be trained in the edge using the shared user data. However, the challenge of maintaining a high QoS is further exacerbated by the fact that the metaverse is characterized by such diverse types of resources amid a highly dynamic demand environment. For example, with the uncertainty in the amount of data users shares, it is difficult for the virtual service provider to subscribe to the correct number of edge servers. Therefore, there exists a crucial need to establish new paradigms of resource allocation that serve not only to meet metaverse users' interests but also to minimize the costs of virtual service providers.

\begin{figure*}[t]
\centering
    \includegraphics[width=14cm, height=4cm,trim={0cm 23.5cm 1cm 0},clip]{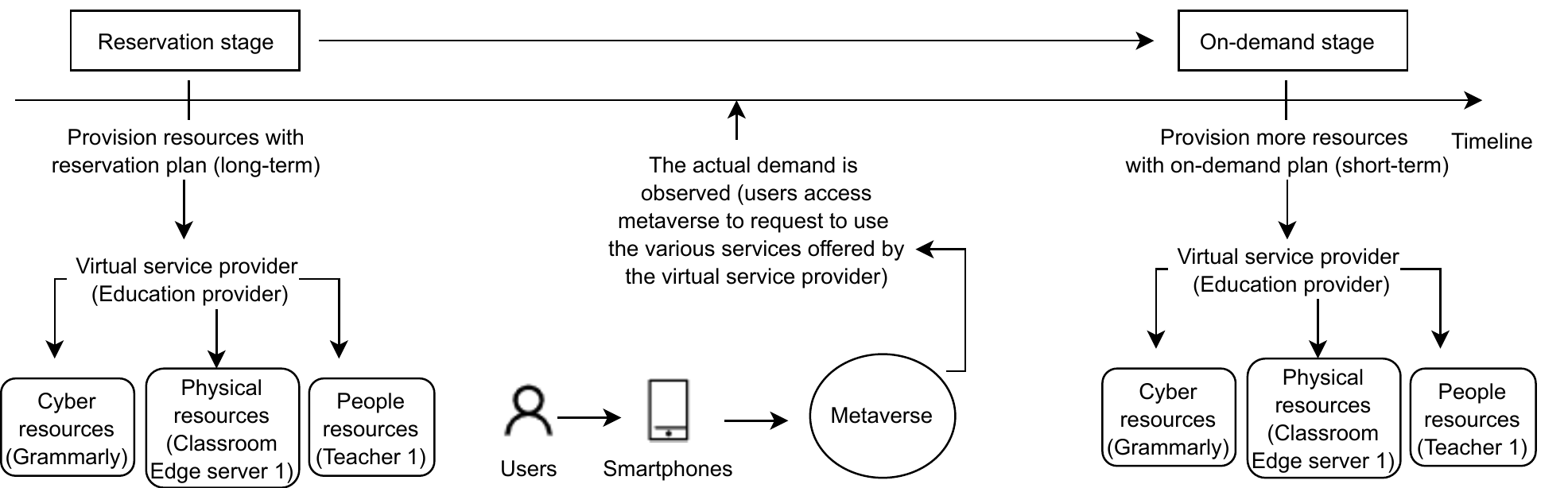}
  \caption{Resource allocation timeline in virtual education case study. $\mathcal{X}=\{x:Grammarly\}$, $\mathcal{Y}=\{y:Classroom\}$, $\mathcal{Z}=\{z:Edge\; server\; 1\}$, and $\mathcal{Y}=\{y:Teacher\; 1\}$.}
  \label{fig:timeline}
  \vspace*{-5mm}
\end{figure*}
In this paper, we propose the case study of  virtual education in the metaverse. The pandemic has necessitated an evolution of the education industry as schools and private enrichment centres face disruption to physical classes \cite{roe2021impact} and more students resort to online resources. The education providers can put up the services that they provide in the metaverse, e.g., personalized lessons in the virtual world delivered by the AI tutors (cyber resources) \cite{ndukwe2019machine}, lessons taught by the teachers (people resources), as well as conducive  classrooms that users may book to study and edge servers (physical resources) to store and process the students' relevant information, e.g., to train the AI system for teaching content personalization. With the help of edge servers, the AI tutors can be deployed at the edge.  For example, by offloading/caching some parts of data and services to reduce latency and to maintain the privacy of user's data as the data is kept at the edge server locally, e.g., servers in a community or the neighborhood area network.

\subsection{Contributions of This Work}
Similar to models in the sharing economy \cite{puschmann2016sharing}, the cyber, physical, and people resources are not always owned by the virtual education providers. Instead, separate entities may own the resources, and the virtual education providers have to subscribe to these resources before offering them to the users. In general, there are two subscription plans i.e., reservation (i.e., long-term) and on-demand plan (i.e., ad-hoc). For example, full-time teachers can be hired for a long term (i.e., considered to be ``reservation'') while part-time teachers may be hired for a few weeks (i.e., considered to be ``on-demand"). Generally, the reservation plan is cheaper than the on-demand plan~\cite{5394134}. Figure~\ref{fig:timeline} shows the subscription timeline of the virtual service providers. The users can access the metaverse using the smartphone interface. Then, the virtual service provider can subscribe to these resources to offer services to the users based on users' demands. However, the virtual service provider will need to decide on the resources to be allocated via reservation plan before an actual student demand is known. Therefore, a resource over-provisioning problem can occur if the virtual service provider subscribes too many resources, i.e., the resources subscribed in the reservation plan are more than the users' demands. In contrast, a resource under-provisioning problem can happen if the virtual service provider subscribes too little resources, i.e., the resources subscribed in the reservation plan are less than the users' demands. Then, the virtual service provider has to meet the demands of the users with the more expensive plan, which is the on-demand plan. Therefore, with the demand uncertainty of the users, we propose a two-stage stochastic integer programming (SIP) for the virtual service providers in metaverse to minimize its operation cost by allocating the resources most strategically.

The contributions of this paper are summarized as follows.
\begin{itemize}
    \item We first provide a brief overview of the metaverse, including the architecture and the metaverse applications. Then, we illustrate with an example of how metaverse is applied to the education sector.
    \item We propose a unified decision-making framework by incorporating the stochastic optimal resource allocation scheme (SORAS) to minimize the cost of the virtual service provider. SORAS uses a two-stage SIP to obtain the optimal decisions of the virtual education provider by minimizing the overall network cost. The formulation incorporates the uncertainty modeling for users' demands.
    \item The performance evaluation substantiates the importance of optimal resource allocation. For example, the performance evaluation shows that intuitively adopting average demand leads to a sub-optimal solution, and it is significantly inferior to the solution from our scheme. 
\end{itemize}

The remainder of the paper is organized as follows: In Section~\ref{system}, we present the system model. In Section~\ref{problem} we formulate the problem. We discuss and analyze the simulation result in Section~\ref{simulation}. Section~\ref{conclusion} concludes the paper.

\section{System Model}\label{system}
We consider the system model from the perspective of one virtual service provider participating in metaverse. As depicted in Fig.~\ref{fig:timeline}, we classified the resources in the metaverse into three major types, i.e., cyber, physical, and people. For simplicity, we did not consider the network resources. These resources can be, for example, used for communications from users to cyber, physical, and people resources.
\begin{itemize}
    \item \textbf{Cyber}: Let $\mathcal{V}~=\{1,\ldots,v,\ldots,V\}$ denote the set of cyber resource and $\mathcal{W}~=\{1,\ldots,w,\ldots,W\}$ denote the set of users in the network. The user $w$ has a demand, i.e., the number of hours to use the cyber resources. Since the cyber resources are implemented on the metaverse platform, the cyber resources can be acquired and deployed in a much shorter time scale than those of physical and people resources.
    \item \textbf{Physical}: Let $\mathcal{X}~=\{1,\ldots,x,\ldots,X\}$ denote the set of physical resources. Edge servers are also part of the physical resources, and it is denoted by a different set notation $\mathcal{Z}~=\{1,\ldots,z,\ldots,Z\}$. Under our consideration, each edge server $z$ can store and process $I_z$ amount of data by the virtual service providers.
    \item \textbf{People}: Let $\mathcal{Y}~=\{1,\ldots,y,\ldots,Y\}$ denote the set of people resources. Virtual service providers hire people resources to provide services to the users. To maximize the service outcome, each people resource can support $E_y$ hours for the users to use. For example, in the virtual education service, teachers are the people resource hired by the virtual education provider to provide consultation to the students (users). People resources are more unreliable (unavailable randomly) than cyber and physical resources. For example, a people resource $y$ can be on medical leave even if $y$ is hired by the virtual service provider.
\end{itemize}

\subsection{Provisioning Plans and costs}
 When subscribing the resources, the virtual service provider can consider the reservation plan or on-demand plan. The virtual service provider can obtain the optimal plan if the provider knows the demand of each user. However, the demand of each user is not always the same. For example, some students may need more hours of consultation with teachers. With the two subscription plans mentioned above, there are two corresponding subscription costs for each of the resources used, i.e., reservation and on-demand costs. The cost function is defined in dollars per resource unit per unit time. $c^r_v$, $c^r_x$, $c^r_z$ and $c^r_y$ are the reservation costs for cyber resource $v$, physical resource $x$, edge server $z$, and people resource $y$ respectively. $c^o_v$, $c^o_x$, $c^o_z$, and $c^o_y$ are the on-demand costs for cyber resource $v$, physical resource $x$, edge server $z$, and people resource $y$ respectively. $c^r_v$ and $c^o_v$ are used to pay for the cyber resources. $c^r_y$ and $c^o_y$ are used to pay for the physical resources. $c^r_y$ and $c^o_y$ are used to pay for the people resources.

\subsection{Uncertainty in Demands}
Under uncertainty of demands, the number of resources required by the user is not exactly known when the reservation of the resources is made. Let $\lambda_i$ denote the $i$ demand scenario of all the users. The set of scenarios is denoted by $\Omega$, i.e., $\lambda_i\in\Omega$. Let $P(\lambda_i)$ denote the probability that scenario $\lambda_i\in\Omega$ is realized, where $P(\lambda_i)$ can be obtained from the historical records~\cite{5394134}. The uncertainty of demands is expressed as follows:

\begin{center}
{\footnotesize
$\lambda_i =\begin{bmatrix}
\arraycolsep=3pt
\medmuskip=1mu 
(F_{1,v},F_{1,x},F_{1,y},\bar{F}_{1,y},F_{1}) & \ldots & (F_{w,v},F_{w,x},F_{w,y},\bar{F}_{w,y},F_{w})
\end{bmatrix}$,}
\end{center}

\noindent where $F_{w,v}$, $F_{w,x}$, and $F_{w,y}$ represent the amounts of time required for cyber resource $v$, physical resource $x$, and people resource $y$ required by user $w$, respectively. $\bar{F}_{w,y}$ is a binary parameter that represents the availability of people resource $y$ when it is requested by the user $w$. $F_{w}$ is a positive parameter that indicates the amount of data in Gb that the user $w$ is willing to share.
If user $w$ has a demand, the user is allocated with a resource duration in each component. For example, $\lambda_i=~\{(F_{w,v}~:~0.3,~F_{w,x}~:~0.3,~F_{w,y}~:~0.4~,~\bar{F}_{w,y}~:1~,~F_{w}~:0.5~)\}$ means that user $w$ requires 0.3hr to the cyber resource $v$, 0.3hr to use physical resource $x$, 0.4hr to use people resource $y$ and people resource $y$ is available as $\bar{F}_{w,y}=1$. User $w$ is also willing to share 0.5Gb of his own data with the virtual service provider. Therefore, $F_{w,v} =0$ $\forall v\in\mathcal{V}\setminus\{v\}$, $F_{w,x}=0$ $\forall x\in\mathcal{X}\setminus\{x\}$, and $F_{w,y}=0$ $\forall y\in\mathcal{Y}\setminus\{y\}$. 

\section{Problem Formulation}\label{problem}
This section introduces the Deterministic Integer Programming (DIP) and Stochastic Integer Programming (SIP) to optimize the resources used by minimizing the virtual service provider total cost.

\subsection{Deterministic Integer Programming}\label{dip}
All the resources can be subscribed to in the reservation plan if users' actual demand is precisely known. Therefore, the on-demand plan is not required. In total, there are four decision variables.
\begin{itemize}
    \item $m_{w,v}^{(\mathrm{c},\mathrm{r})}$ indicates the duration that is reserved for user $w$ to use cyber resource $v$. For example, $m_{w,v}^{(\mathrm{c},\mathrm{r})}=1.2$ means that the virtual education provider reserves 1.2 hours for user $w$ to use cyber resource $v$.
    \item $m_{w,x}^{(\mathrm{p},\mathrm{r})}$ indicates the duration that is reserved for user $w$ to use physical resource $x$. 
    \item $m_{w,z}^{(\mathrm{p},\mathrm{r})}$ indicates whether edge server $z$ is reserved to store and process the data from user $w$. 
    \item $m_{w,y}^{(\mathrm{h},\mathrm{r})}$ indicates the duration that is reserved for user $w$ to use people resource $y$.
\end{itemize}

A DIP can be formulated to minimize the total cost of the virtual education provider as follows:

\noindent $\displaystyle\min_{m_{w,v}^{(\mathrm{c},\mathrm{r})},m_{w,x}^{(\mathrm{p},\mathrm{r})},m_{w,z}^{(\mathrm{p},\mathrm{r})},m_{w,y}^{(\mathrm{h},\mathrm{r})}}$:

\begin{align}\label{dip1}
    \sum_{w\in \mathcal{W}}\biggl(\sum_{v\in\mathcal{V}}(m_{w,v}^{(\mathrm{c},\mathrm{r})}c^r_v)+ \sum_{x\in\mathcal{X}}(m_{w,x}^{(\mathrm{p},\mathrm{r})}c^r_x)+\nonumber\\\sum_{z\in\mathcal{Z}}m_{w,z}^{(\mathrm{p},\mathrm{r})}c^r_z+\sum_{y\in\mathcal{Y}}m_{w,y}^{(\mathrm{h},\mathrm{r})}c^r_y\biggl),
\end{align}
subject to: 
\begin{align}
    &m_{w,v}^{(\mathrm{c},\mathrm{r})} = D_{w,v}, &\forall w\in \mathcal{W}, \forall v\in\mathcal{V},\label{dip_cons1}\\
    &m_{w,x}^{(\mathrm{p},\mathrm{r})} = D_{w,x}, &\forall w\in \mathcal{W}, \forall x\in\mathcal{X},\label{dip_cons2}\\
    &\bar{D}_{w,y}m_{w,y}^{(\mathrm{h},\mathrm{r})} = D_{w,y}, &\forall y\in\mathcal{Y}, \forall w\in \mathcal{W},\label{dip_cons3}\\
    &\sum_{w\in \mathcal{W}}m_{w,y}^{(\mathrm{h},\mathrm{r})}\leq E_y, &\forall y\in\mathcal{Y},\label{dip_cons4}\\
    &\sum_{z\in\mathcal{Z}}m_{w,z}^{(\mathrm{p},\mathrm{r})}I_z\geq D_w, &\forall w\in\mathcal{W},\label{dip_cons5}\\
    &\sum_{w\in\mathcal{W}}m_{w,z}^{(\mathrm{p},\mathrm{r})} \leq 1, &\forall z\in\mathcal{Z},\label{dip_cons6}\\
    &m_{w,z}^{(\mathrm{p},\mathrm{r})} \in\{0,1\}, &\forall z\in\mathcal{Z},\label{dip_cons7}\\
    &m_{w,y}^{(\mathrm{h},\mathrm{r})} \in\{0,\ldots,E_y\},&\forall w\in \mathcal{W},\forall y\in\mathcal{Y},\label{dip_cons8}\\
    &m_{w,v}^{(\mathrm{c},\mathrm{r})}, m_{w,x}^{(\mathrm{p},\mathrm{r})}, \in\mathbb{Z}^+, &\forall w\in \mathcal{W}, \forall v\in\mathcal{V},\forall x\in\mathcal{X}.\label{dip_cons9}
\end{align}
The objective function in~(\ref{dip1}) is to minimize the total cost due to resource reservation. $D_{w,v}$, $D_{w,x}$, and $D_{w,y}$ are the actual time demands for user $w$ to use cyber resource $v$, physical resource $x$, and people resource $y$. $\bar{D}_{w,y}$ is the actual availability of people resource $y$ when it is requested by user $w$. $D_w$ is the actual amount of data that user $w$ shared. The constraints in (\ref{dip_cons1})~-~(\ref{dip_cons3}) ensure that the demand is met. In~(\ref{dip_cons3}), if people resource $y$ is not available, user $w$ should request the service from another available people resource. (\ref{dip_cons4}) ensures that the number of hours allocated to people resource $y$ does not exceed $E_y$. (\ref{dip_cons5}) ensures that the number of edge servers should be large enough to support the amount of data user $w$ shared. (\ref{dip_cons6}) ensures that each edge server can only be used once in the whole network. (\ref{dip_cons7}) indicates that $m_{w,z}^{(\mathrm{p},\mathrm{r})}$ is a binary variable. (\ref{dip_cons8}) and~(\ref{dip_cons9}) indicate that $m_{w,v}^{(\mathrm{c},\mathrm{r})}$, $m_{w,x}^{(\mathrm{p},\mathrm{r})}$, and $m_{w,y}^{(\mathrm{h},\mathrm{r})}$ are positive decision variables.
\subsection{Stochastic Integer Programming}\label{sip_pro}
If the demands for the resources are not known, the DIP formulated in~(\ref{dip1})~-~(\ref{dip_cons4}) is no longer applicable. Therefore, SIP with a two-stage recourse is developed. This section introduces the SIP to minimize the total cost of the network by optimizing the number of hours allocated to each user for different types of resources. The first stage consists of all decisions that must be selected before the demands are realized and observed. The virtual service provider has to subscribe to the duration for the resources to be used in advance before observing the demands. In the second stage, decisions are allowed to adapt to the demand observed. After the demand is observed, the virtual service provider has to pay for the additional hours needed if the reserved duration is shorter than the demand.

Other than the four decision variables listed in Section~\ref{dip}, there are four more decision variables in the SIP formulation.
\begin{itemize}
    \item $m_{w,v}^{(\mathrm{c},\mathrm{o})}(\lambda_i)$ indicates the duration that is used as on-demand for user $w$ to use cyber resource $v$ in scenario $\lambda_i$.
    \item $m_{w,x}^{(\mathrm{p},\mathrm{o})}(\lambda_i)$ indicates the duration that is used as on-demand for user $w$ to use physical resource $x$ in scenario $\lambda_i$.
    \item $m_{w,z}^{(\mathrm{p},\mathrm{o})}(\lambda_i)$ indicates if edge server $z$ is used as on-demand to store and process the data shared by user $w$ in scenario $\lambda_i$.
    \item $m_{w}^{(\mathrm{h},\mathrm{o})}(\lambda_i)$ indicates the duration that the virtual education provider has to outsource for user $w$ for people resource in scenario $\lambda_i$.
\end{itemize}

The objective function given in~(\ref{sip1}) and~(\ref{sip2}) is to minimize the cost of the resource allocation. The expressions in~(\ref{sip1}) and~(\ref{sip2}) represent the first- and second-stage SIP, respectively. The SIP formulation can be expressed as follows:
\noindent $\displaystyle\min_{m_{w,v}^{(\mathrm{c},\mathrm{r})},\ldots,m^{(\mathrm{h},\mathrm{o})}_{w,y}(\lambda_i)}$:

 \begin{align}\label{sip1}
    \sum_{w\in \mathcal{W}}\biggl(\sum_{v\in\mathcal{V}}(m_{w,v}^{(\mathrm{c},\mathrm{r})}c^r_v)+ \sum_{x\in\mathcal{X}}(m^{(\mathrm{p},\mathrm{r})}_{w,x}c^r_x)+\sum_{z\in\mathcal{Z}}m_{w,z}^{(\mathrm{p},\mathrm{r})}c^r_z+\nonumber\\\sum_{y\in\mathcal{Y}}m^{(\mathrm{h},\mathrm{r})}_{w,y}c^r_y\biggl)+\mathbb{E}\biggl[\mathcal{Q}(m_{w,v}^{(\mathrm{c},\mathrm{r})},m^{(\mathrm{p},\mathrm{r})}_{w,x},m_{w,z}^{(\mathrm{p},\mathrm{r})},m^{(\mathrm{h},\mathrm{r})}_{w,y},\lambda_i)\biggr],
\end{align}
where
\begin{align}\label{sip2}
    \mathcal{Q}(m_{w,v}^{(\mathrm{c},\mathrm{r})},m^{(\mathrm{p},\mathrm{r})}_{w,x},m_{w,z}^{(\mathrm{p},\mathrm{r})},m^{(\mathrm{h},\mathrm{r})}_{w,y},\lambda_i) =\sum_{\lambda_i\in\Omega}P(\lambda_i)\sum_{w\in \mathcal{W}}\nonumber\\\biggl(\sum_{v\in\mathcal{V}}
    (m_{w,v}^{(\mathrm{c},\mathrm{o})}(\lambda_i)c^o_v)+ \sum_{x\in\mathcal{X}}(m^{(\mathrm{p},\mathrm{o})}_{w,x}(\lambda_i)c^o_x)+\nonumber\\\sum_{z\in\mathcal{Z}}m_{w,z}^{(\mathrm{p},\mathrm{o})}(\lambda_i)c^o_z+
    \sum_{y\in\mathcal{Y}}m^{(\mathrm{h},\mathrm{o})}_{w,y}(\lambda_i)c^o_y\biggl),
\end{align}
subject to: 
\begin{align}
    m^{(\mathrm{c},\mathrm{r})}_{w,v} + m^{(\mathrm{c},\mathrm{o})}_{w,v}(\lambda_i) \geq F_{w,v}(\lambda_i),\hspace*{+25mm}\nonumber\\ \forall w\in \mathcal{W}, \forall v\in\mathcal{V}, \forall \lambda_i\in\Omega,\label{sip_cons1}
\end{align}
\begin{align}
    m^{(\mathrm{p},\mathrm{r})}_{w,x} + m^{(\mathrm{p},\mathrm{o})}_{w,x}(\lambda_i) \geq F_{w,x}(\lambda_i),\hspace*{+25mm}\nonumber\\ \forall w\in \mathcal{W}, \forall x\in\mathcal{X},\forall \lambda_i\in\Omega,\label{sip_cons2}
\end{align}
\begin{align}
    \bar{F}_{w,y}(\lambda_i)m^{(\mathrm{h},\mathrm{r})}_{w,y} + m^{(\mathrm{h},\mathrm{o})}_{w}(\lambda_i) \geq F_{w,y}(\lambda_i),\hspace*{+11mm}\nonumber\\ \forall y\in\mathcal{Y}, \forall w\in \mathcal{W},\forall \lambda_i\in\Omega,\hspace*{-1mm}\label{sip_cons3}
\end{align}
\begin{align}
    &\sum_{w\in \mathcal{W}}m^{(\mathrm{h},\mathrm{r})}_{w,y}\leq E_y,\hspace*{+18mm} &\forall y\in\mathcal{Y},\label{sip_cons4}\\
    &\sum_{w\in\mathcal{W}}m_{w,z}^{(\mathrm{p},\mathrm{r})} \leq 1, &\forall z\in\mathcal{Z},\label{sip_cons6}\\
    &\sum_{w\in\mathcal{W}}m_{w,z}^{(\mathrm{p},\mathrm{r})}(\lambda_i) \leq 1, &\forall z\in\mathcal{Z},\forall \lambda_i\in\Omega,\label{sip_cons7}
\end{align}
\begin{align}
    \sum_{z\in\mathcal{Z}}m_{w,z}^{(\mathrm{p},\mathrm{r})}I_z +\sum_{z\in\mathcal{Z}}m_{w,z}^{(\mathrm{p},\mathrm{o})}(\lambda_i)I_z \geq F_w(\lambda_i),\hspace*{+8mm} \nonumber\\ \forall w\in \mathcal{W},\forall \lambda_i\in\Omega,\hspace*{-0mm}\label{sip_cons8}
\end{align}
\begin{align}
    &m_{w,z}^{(\mathrm{p},\mathrm{r})}+ m_{w,z}^{(\mathrm{p},\mathrm{r})}(\lambda_i) =1, &\forall w\in \mathcal{W},\forall z\in\mathcal{Z},\forall \lambda_i\in\Omega,\label{sip_cons9}\\
    &m_{w,z}^{(\mathrm{p},\mathrm{r})},m_{w,z}^{(\mathrm{p},\mathrm{r})}(\lambda_i) \in\{0,1\}, &\forall w\in \mathcal{W},\forall z\in\mathcal{Z},\forall \lambda_i\in\Omega,\label{sip_cons10}
\end{align}
\begin{align}
    m^{(\mathrm{h},\mathrm{r})}_{w,y},m^{(\mathrm{h},\mathrm{o})}_{w,y}(\lambda_i)\in\{0,\ldots,E_y\},\hspace*{+28mm}\nonumber\\\forall w\in \mathcal{W},\forall y\in\mathcal{Y},\forall \lambda_i\in\Omega,\label{sip_cons11}
\end{align}
\begin{align}
    m^{(\mathrm{c},\mathrm{r})}_{w,v}, m^{(\mathrm{p},\mathrm{r})}_{w,x}, m^{(\mathrm{c},\mathrm{o})}_{w,v}(\lambda_i),m^{(\mathrm{p},\mathrm{o})}_{w,x}(\lambda_i) \in\mathbb{Z}^+,\hspace*{+8mm} \nonumber\\\forall w\in \mathcal{W}, \forall v\in\mathcal{V},\forall x\in\mathcal{X},\forall \lambda_i\in\Omega.\label{sip_cons12}
\end{align}
(\ref{sip_cons1})~-~(\ref{sip_cons3}) ensure that each user's demand has to be met by using the reservation and on-demand plan. In~(\ref{sip_cons3}), if the people resource $y$ is not available, user $w$ should request the on-demand service from another available people resource. (\ref{sip_cons4}) ensures that the number of hours allocated to people resource $y$ in the reservation plan does not exceed $E_y$. (\ref{sip_cons6}) and~(\ref{sip_cons7}) ensure that each edge server can only be used once in the whole network. (\ref{sip_cons8}) ensures that there should be enough edge servers for the virtual service provider to store and process the shared data. (\ref{sip_cons9}) ensures that each edge server can subscribe only one time in each of the plans. (\ref{sip_cons10}) indicates that $m^{(\mathrm{h},\mathrm{r})}_{w,y}$ and $m^{(\mathrm{h},\mathrm{o})}_{w,y}(\lambda_i)$ are binary variables. (\ref{sip_cons11}) and~(\ref{sip_cons12}) indicate that $m^{(\mathrm{c},\mathrm{r})}_{w,v}$, $m^{(\mathrm{p},\mathrm{r})}_{w,x}$, $m^{(\mathrm{h},\mathrm{r})}_{w,y}$, $m^{(\mathrm{c},\mathrm{o})}_{w,v}(\lambda_i)$, $m^{(\mathrm{p},\mathrm{o})}_{w,x}(\lambda_i)$, and $m^{(\mathrm{h},\mathrm{o})}_{w,y}(\lambda_i)$ are positive decision variables.

\begin{figure}[t]
    \includegraphics[width=8cm, height=4cm,trim={0cm 9cm 0cm 0},clip]{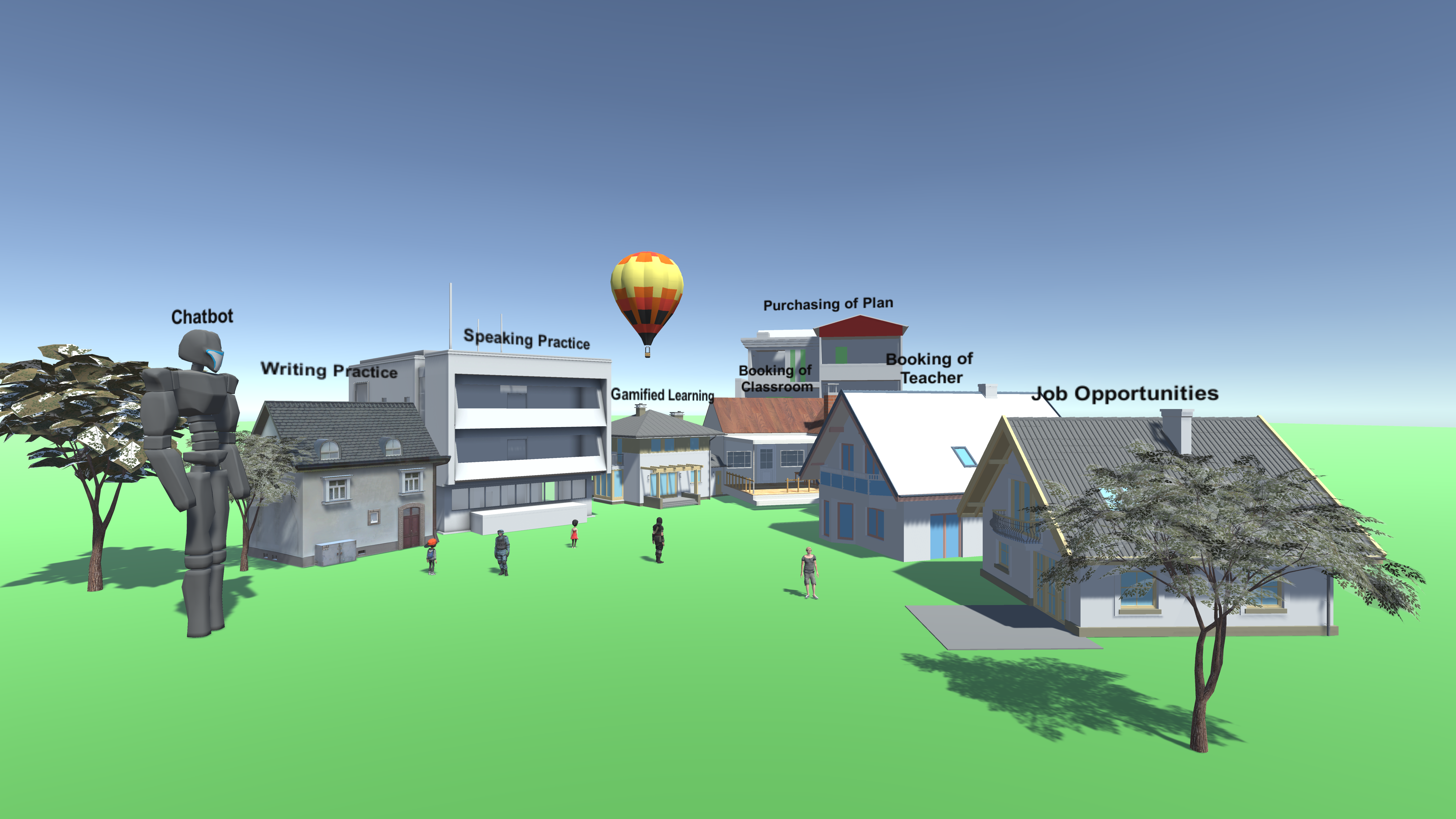}
  \caption{3D English Education Metaverse prototype through metaverse viewer.}
  \label{fig:prototype}
\end{figure}
\begin{figure*}[t]
\centering
\begin{multicols}{4}
  \includegraphics[width=0.95\columnwidth]{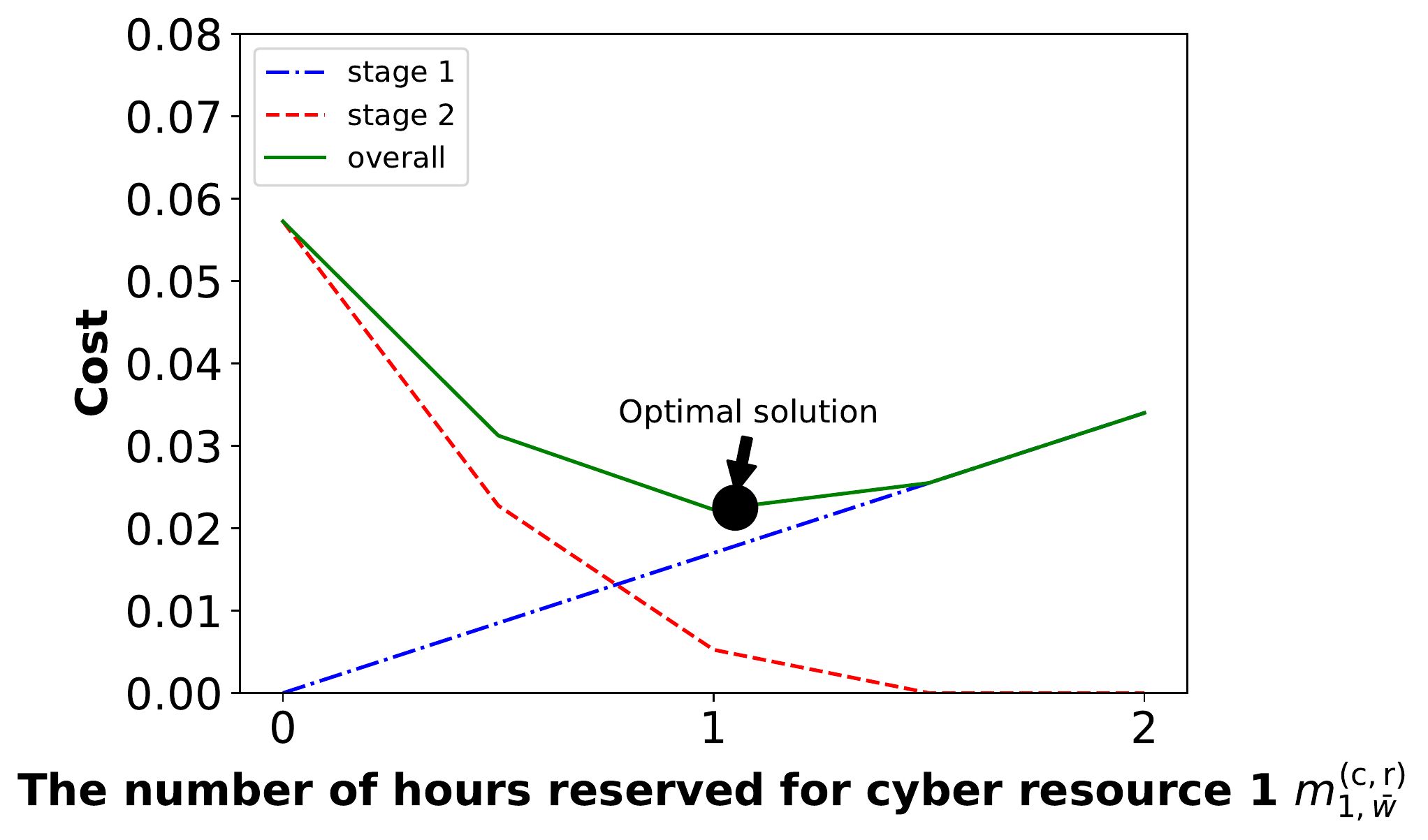}
  \caption{The cost structure in a simple SIP network for using cyber resource $\bar{w}$.}
  \label{fig:optimal cost cyber}
  \includegraphics[width=0.95\columnwidth]{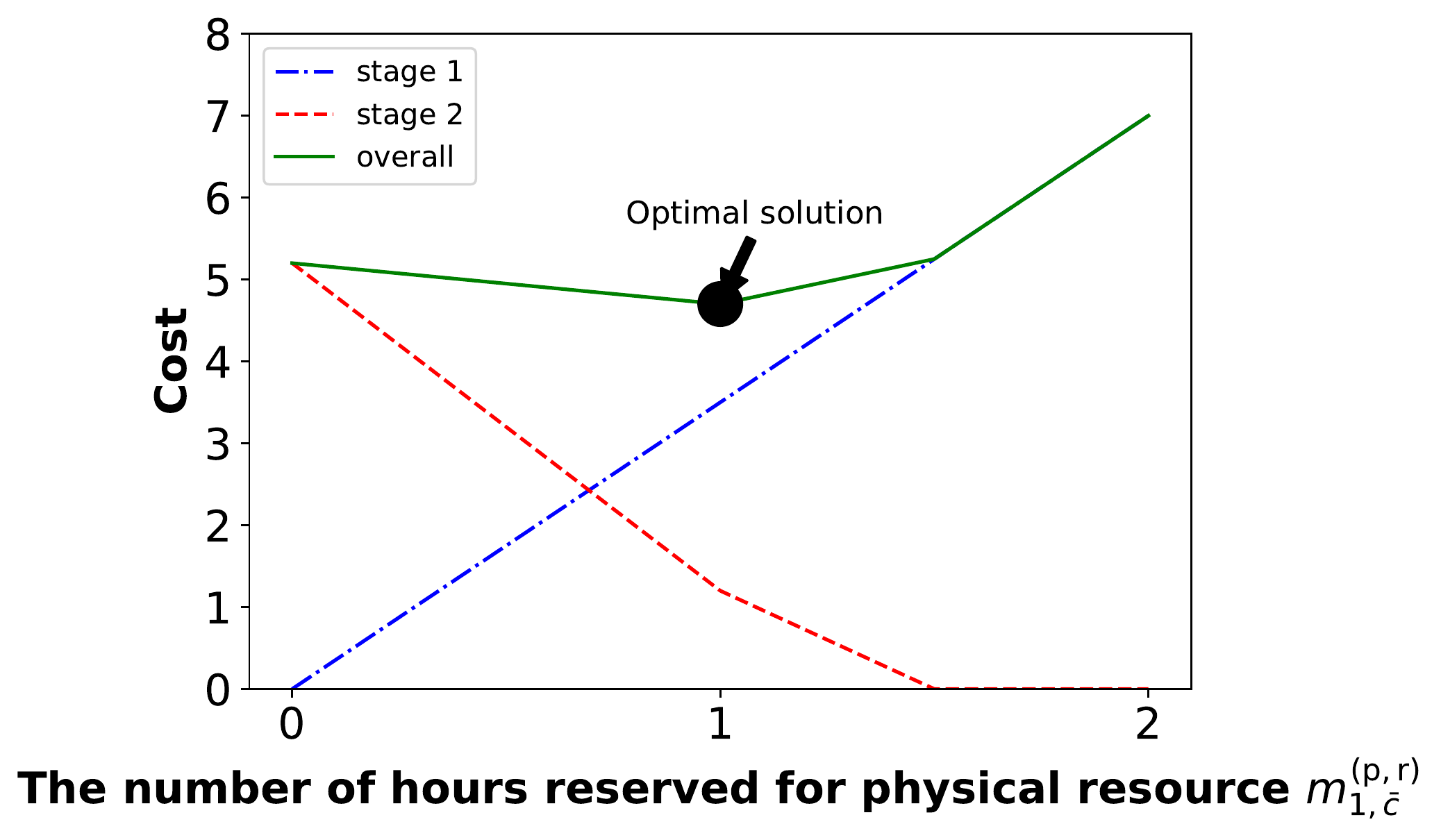}\par
  \caption{The cost structure in a simple SIP network for using physical resource $\bar{c}$.}
  \label{fig:optimal cost physical}
  \includegraphics[width=0.85\columnwidth]{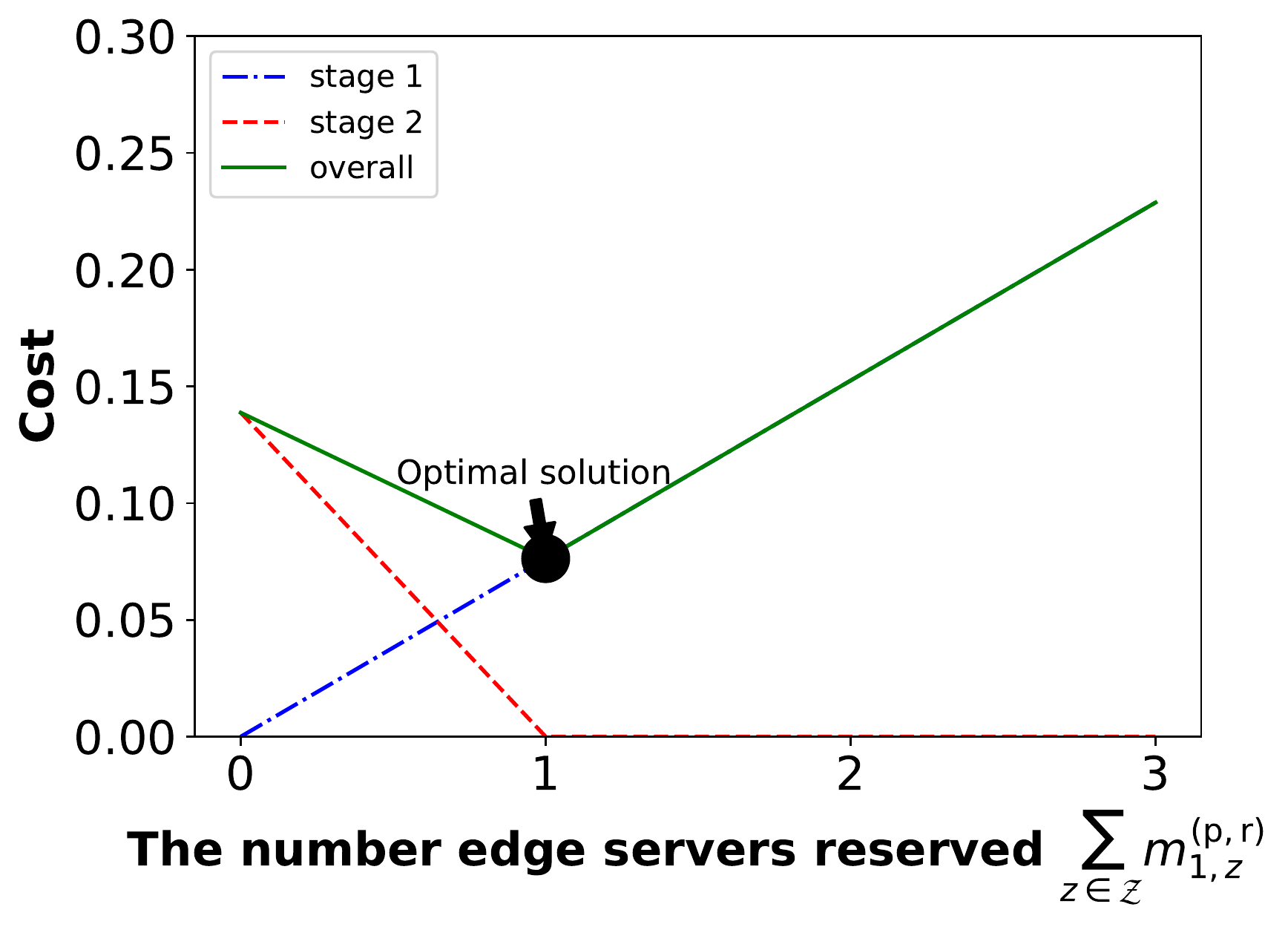}\par
  \caption{The cost structure in a simple SIP network for using edge servers.}
  \label{fig:optimal cost edge}
  \includegraphics[width=0.95\columnwidth]{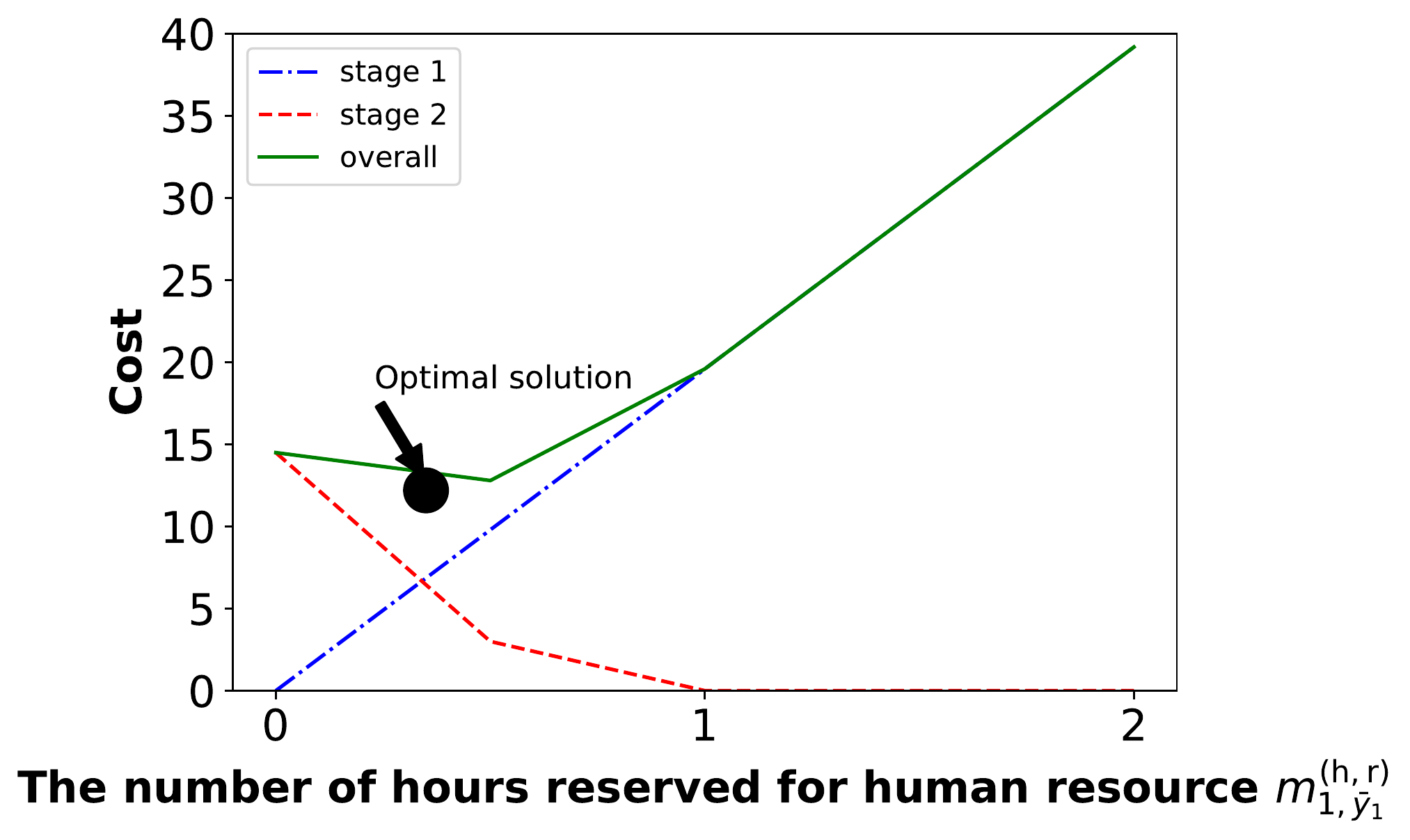}\par
  \caption{The cost structure in a simple SIP network for using people resource $\bar{y}_1$.}
  \label{fig:optimal cost human}
\end{multicols}
\vspace*{-5mm}
\end{figure*}

\section{Performance Evaluation}\label{simulation}
\subsection{English Education Metaverse Prototype}
We develop the 3D English Education metaverse (EEM) on Unity~\cite{unity}, a cross-platform development engine. As shown in Fig.~\ref{fig:prototype}, we illustrate the interface of the proposed 3D English Education Metaverse prototype through the implemented metaverse viewer. The metaverse viewer is built for the users to interact with the metaverse. By using the smartphone as the platform, the users can access the metaverse any time and any place with Internet access. 

In this EEM, the resource owners first charge the virtual education provider with the on-demand (hourly) or the reservation cost (per semester). Then, after the resources are purchased, the virtual education provider offers the users services according to the users' demands. Then, the users can navigate about in the virtual world using their avatars to choose the service they want. All the users can communicate with the NPCs if any of them has any issues. Each user has different demands, and the virtual education provider needs to allocate the optimal resource to serve the users. The two-stage SIP from Section~\ref{sip_pro} is used to optimize resource allocation by minimizing each user's reservation and on-demand cost. The first stage (\ref{sip1}) is to minimize the reservation cost, and the second stage (\ref{sip2}) is to minimize the on-demand cost.

\begin{figure*}[t]
\centering
\begin{multicols}{4}
  \includegraphics[width=0.95\columnwidth]{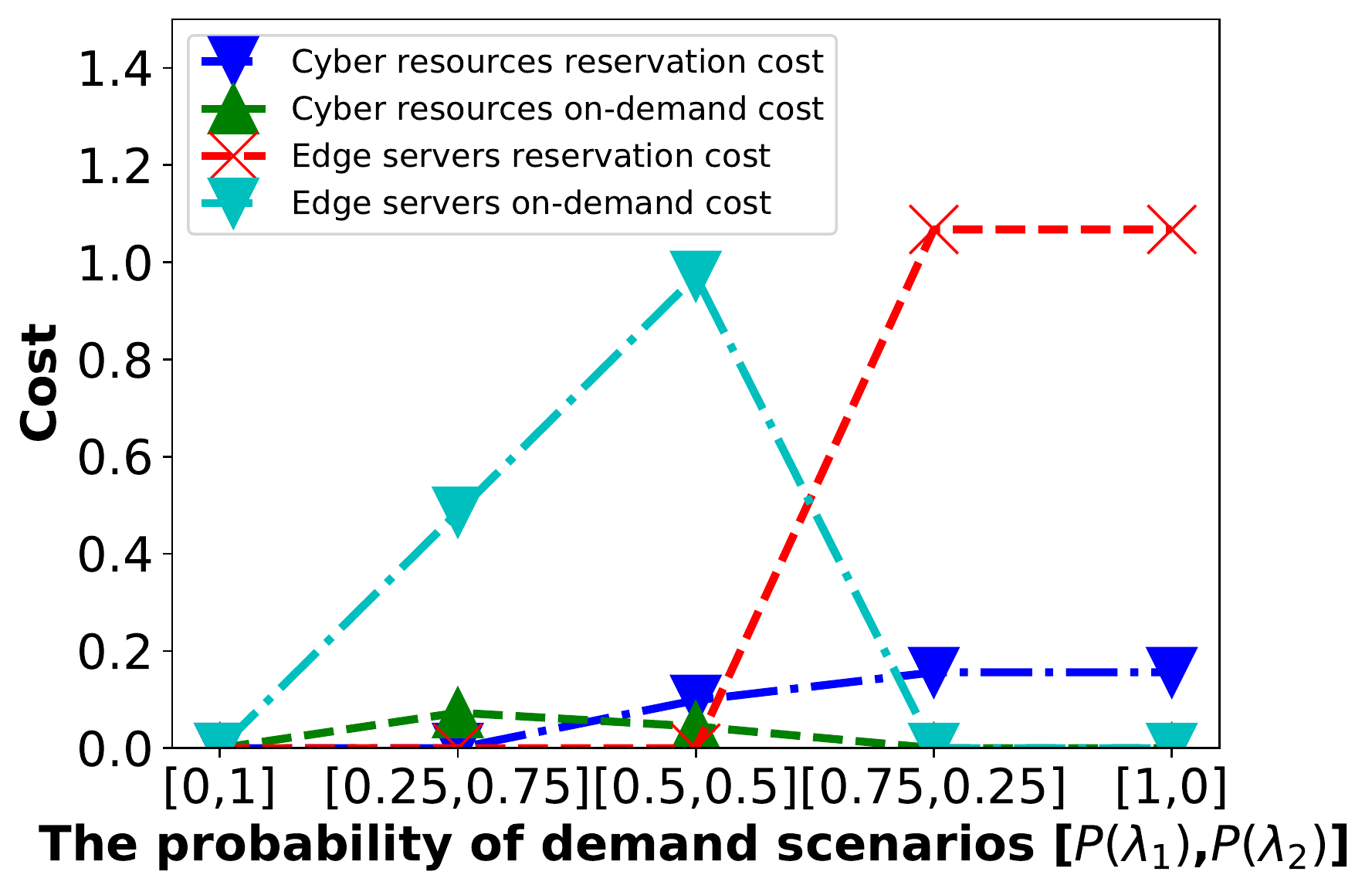}\par
  \caption{The cost of edge servers and cyber resources.}
  \label{fig:smallscale}
  \includegraphics[width=0.95\columnwidth]{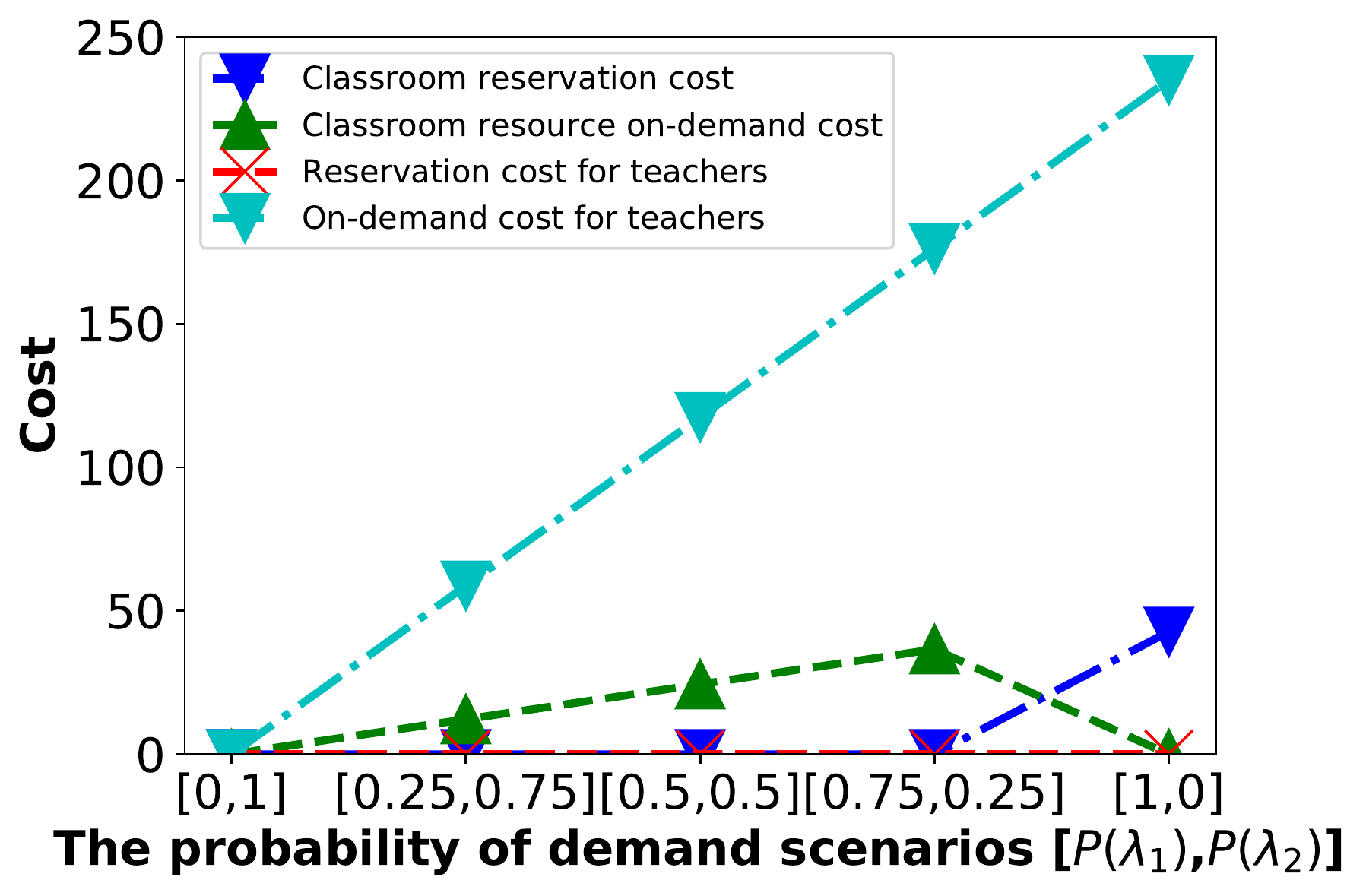}
  \caption{The cost of physical and people resources.}
  \label{fig:largescale}
    \includegraphics[width=0.85\columnwidth]{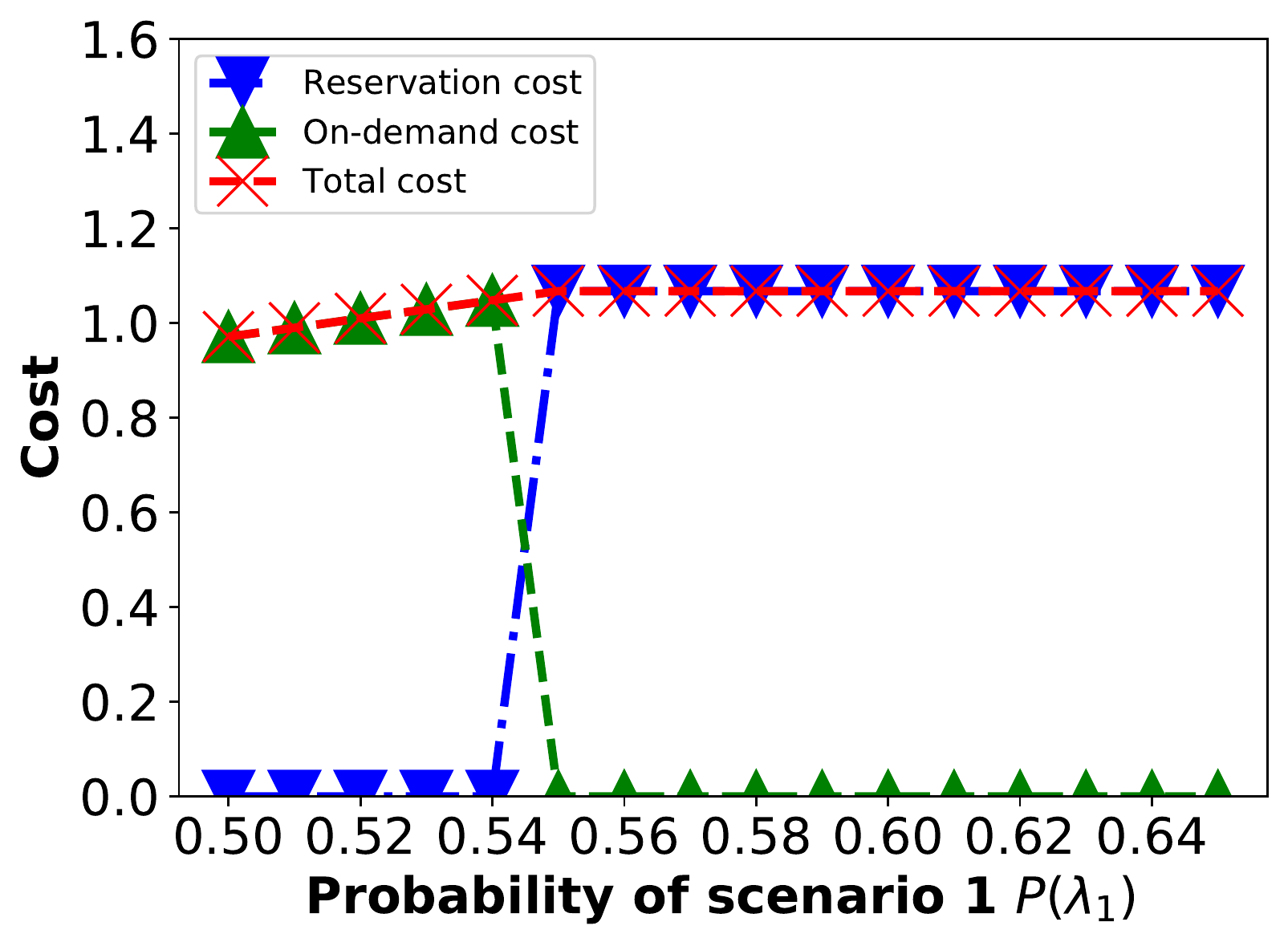}\par
  \caption{The probability threshold of edge servers.}
  \label{fig:threshold}
 \includegraphics[width=0.95\columnwidth]{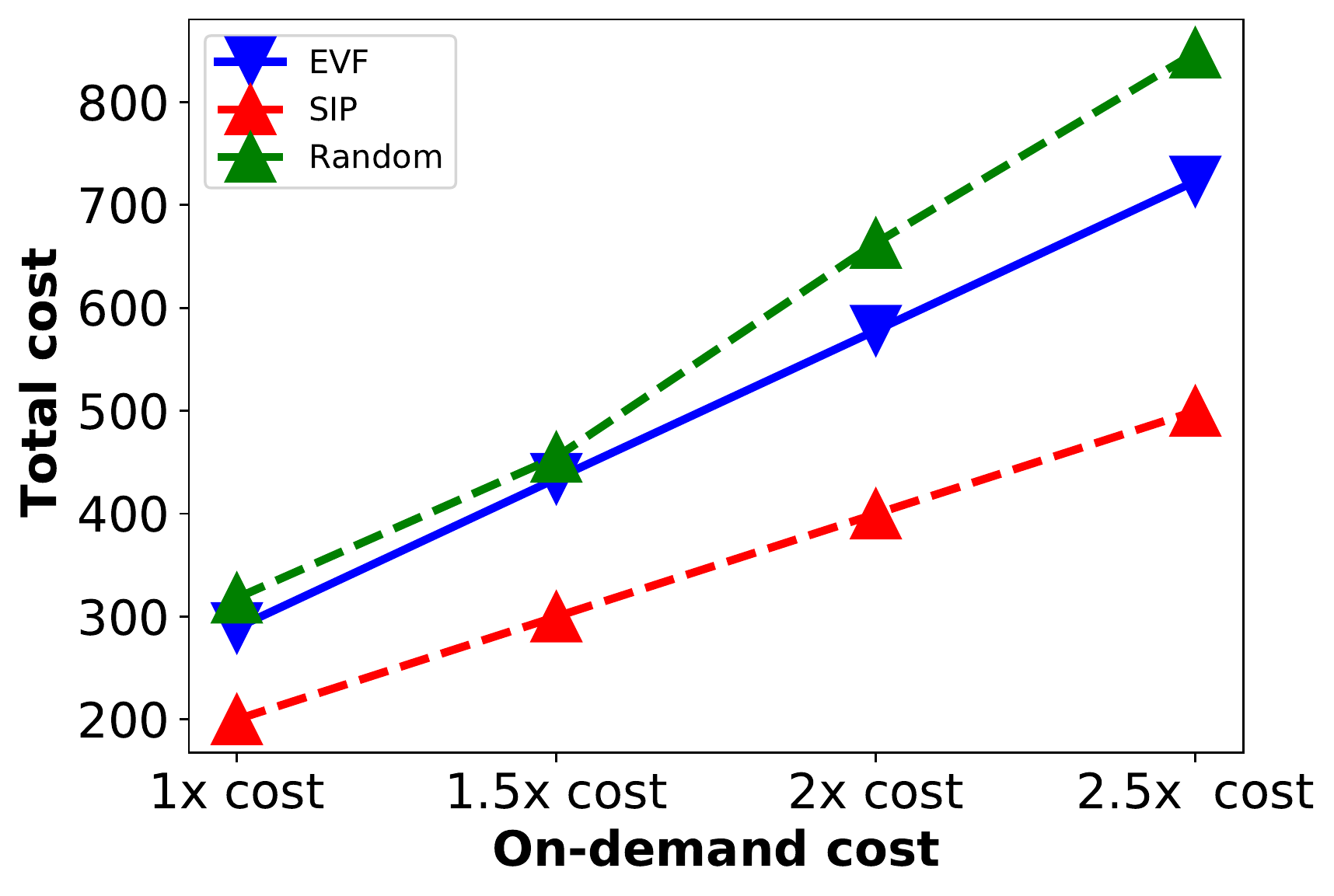}\par
  \caption{SIP comparing with EVF and randoms cheme.}
  \label{fig:evf}
\end{multicols}
\vspace*{-5mm}
\end{figure*}

\subsection{Parameter Setting}
We consider the English language school to be is the virtual service provider that offers services to ten users, and there are 20 edge servers available $|\mathcal{Z}|=20$. Teachers are the people resources $\mathcal{Y}~=~\{\bar{y}_1,\bar{y}_2,\bar{y}_3,\bar{y}_4\}$. $\mathcal{V}~=~\{\bar{w},\bar{s},\bar{l}\}$, where $\bar{w}$ represents the writing practice such as Grammarly~\cite{grammarly}, $\bar{s}$ represents the speaking practice such as ELSA~\cite{ELSA}, and $\bar{l}$ represents gamified learning such as Brainscape~\cite{brainscape}. $\mathcal{X}=\{\bar{c}\}$, where $\bar{c}$ represents a physical classroom.

All the resource owners offer a short-term plan and a long-term plan. We consider the long-term plan as the reservation plan, and the short-term plan is the on-demand plan. Therefore, $c^r_{v:\bar{w}}=\$0.017$/hr, $c^o_{v:\bar{w}}=\$0.035$/hr, $c^r_{v:\bar{s}}=\$0.005$/hr, $c^o_{v:\bar{s}}=\$0.009$/hr, $c^r_{v:\bar{l}}=\$0.010$/hr, $c^o_{v:\bar{l}}=\$0.014$/hr, $c^r_{x:\bar{c}}=\$3.5$/hr, and $c^o_{x:\bar{c}}=\$4$/hr~\cite{grammarly,ELSA,brainscape,classroom}. We assume that the virtual education provider update its data every hour by store and process the data that the users are willing to share. Thus, the costs are $c^r_z=\$0.07625$/use and $c^o_z= \$0.13875$/use~\cite{smartphone}. 
We use the full-time teacher's salary as the reservation plan and the part-time teacher's salary as the on-demand plan. Then, $c^o_{y}=\$19.6$/hr and $c^r_{y}=\$25$/hr~\cite{teacher}.

To solve SIP, we assume that the probability distribution of all scenarios in set $\Omega$ are known~\cite{dyer2006computational}, then, the complexity of the problem depends on the total number of scenarios in stage two~\cite{dyer2006computational}. For example in Section~\ref{sip_pro}, the complexity for the formulated two-stage SIP is $|\Omega|$. For the presented experiments, we implement the SIP model using GAMS script~\cite{chattopadhyay1999application}.

\subsection{Simulations}
\subsubsection{Cost structure} We first study the cost structure of the network. As an illustration, a simple network is considered with only one component in each resource set, e.g., $|\mathcal{V}|=1$, one user, and two demand scenarios $|\Omega|=2$. The first demand scenario is $\lambda_1$, the user has a demand to use all the resources and is willing to share some data. The second demand scenario is $\lambda_2$, and the user has another demand to use all the resources and is willing to share a different amount of data. We consider a stochastic system with $P(\lambda_1)= 0.6$ and $P(\lambda_1)= 0.4$. We observe the cost structure of the network by varying the number of hours reserved for all the resources except for edge servers, where we vary the number of edge servers reserved. The cost structures are shown in Figs.~\ref{fig:optimal cost cyber}~-~\ref{fig:optimal cost human}. In Fig.~\ref{fig:optimal cost cyber}, the costs in the first and second stages, as well as the overall cost under different number of hours reserved for cyber resource $\bar{w}$ is presented. We can observe that the first stage cost (reservation cost), increases as the number of hours of resource reserved increases. With more hours reserved in the first stage, stage 2 cost is reduced as the need for on-demand reduces. It can be identified that even in this simple network, the optimal solution is not trivial to obtain due to the uncertainty of demands. For example, the optimal cost is not the point where the cost in the first and second stages intersect. Therefore, SIP formulation is required to guarantee the minimum cost to the network.


\subsubsection{Individual resources}\label{setup2} There are two demand scenarios $|\Omega|=2$, including i) all the users have demands and all the teachers are not available, e.g., medical leave, denoted by $\lambda_1$ and ii) all the users have no demand denoted by $\lambda_2$. We analyze individually each resource by varying both the demand probabilities $P(\lambda_1)$ and $P(\lambda_2)$. The cost of resources are shown in Figs.~\ref{fig:smallscale} and~\ref{fig:largescale}. For better visualization, the cost of the resources is split into two parts. When the probability of the demand $P(\lambda_1)$ is low, it is cheaper for the virtual education provider to subscribe to the resources using the on-demand plan. From the results, the decision on choosing the subscription plans is also affected by the magnitude of differences between the reservation cost and on-demand cost. If the reservation cost is close to the on-demand cost, the virtual education provider will only subscribe to that resource when $P(\lambda_1)$ is very high. For example, the virtual education provider only books the classroom using the reservation plan when $P(\lambda_1)=1$. However, if the reservation cost is much lower than the on-demand cost, e.g., $c^o_{v:1}$ is more than two times $c^r_{v:1}$, it is cheaper for the virtual education provider to subscribe to that resource in the reservation plan when $P(\lambda_1)$ is not high.

\subsubsection{Probability threshold of demand scenario} We consider the settings similar to Section \ref{setup2}. From Fig.~\ref{fig:smallscale}, the virtual education provider changes its decisions for edge servers when $P(\lambda_1)$ increases from 0.5 to 0.75. In this simulation, we will increase $P(\lambda_1)$ with a small step of 0.01 to determine the probability threshold that causes the virtual education provider to change subscription decisions. The result is illustrated in Fig.~\ref{fig:threshold}. When the probability of scenario 1 is greater or equal to 0.55, we can observe the probability threshold for the virtual education provider to consider the reservation plan as the cheaper plan than that of the on-demand plan.

\subsubsection{Comparing between EVF, SIP and random scheme}
We compare the SIP with expected-value formulation (EVF)~\cite{5394134} as well as the random scheme. For EVF, the number of hours in the first stage is fixed by the average value of demand, an approximation scheme. In the random scheme, the values of the decision variables are randomly generated. We vary the on-demand cost to compare the difference between EVF, SIP, and random schemes. Fig.~\ref{fig:evf} depicts the comparison result. As shown in the result, EVF and random scheme cannot adapt to the change in cost. On the other hand, SIP can always achieve the best solution among the three to reduce the on-demand cost.

\section{Conclusion}\label{conclusion}
In this paper, we have presented a unified resource allocation framework SORAS for the edge-based metaverse in a case study of education sector. To achieve the optimal allocation, SORAS minimizes the total cost of the network. The performance evaluation of SORAS has been performed by numerical studies and simulations.  Compared to the benchmark, SORAS based on SIP can achieve the best solution as it can better adapt to changes in the probability of users' demands.
\bibliographystyle{IEEEtran}
\bibliography{mybibliography.bib}
\end{document}